\documentclass[showpacs, preprintnumbers, prl, superscriptaddress, floatfix, twocolumn, longbibliography]{revtex4-1}
\usepackage{amssymb}
\usepackage{amsmath}
\usepackage{float}
\usepackage{graphicx}
\usepackage{epsfig}
\usepackage[T1]{fontenc}
\usepackage{color}
\usepackage{verbatim}

\usepackage{soul}
\setstcolor{red}
\setulcolor{red}
\usepackage[colorlinks,linkcolor=blue,hyperindex,CJKbookmarks]{hyperref}

\usepackage[utf8]{inputenc}

\newcommand{\beq}{\begin{equation}}
\newcommand{\eeq}{\end{equation}}
\newcommand{\bea}{\begin{eqnarray}}
\newcommand{\eea}{\end{eqnarray}}

\definecolor{darkgreen}{rgb}{0, 0.4, 0}

\def\MnOTF{MnBi$_{\text{2}}$Te$_{\text{4}}$}
\def\MnOFS{MnBi$_{\text{4}}$Te$_{\text{7}}$}

\newcommand{\MBT}{(MnBi\textsubscript{2}Te\textsubscript{4})(Bi\textsubscript{2}Te\textsubscript{3})\textsubscript{n}}
\newcommand{\HIIc}{$H \parallel c$}
\newcommand{\HIIab}{$H \parallel ab$}
\newcommand{\mbt} {\MnOTF}
\newcommand{\mbtl} {\MnOFS}

\newcommand{\corr}[1]{\textcolor{black}{#1}}

\definecolor{Ora}{cmyk}{0, 0.6, 0.8, 0.4}
\definecolor{Ora2}{cmyk}{0, 0.2, 0.8, 0.4}
\definecolor{violet}{rgb}{0.53, 0.0, 0.69}

\begin{document}

\title{Strongly anisotropic spin dynamics in magnetic topological insulators}

\author{A.~Alfonsov}
\thanks{These authors contributed equally to this work.}
\affiliation{Leibniz IFW Dresden, D-01069 Dresden, Germany}
\author{J.~I.~Facio}
\thanks{These authors contributed equally to this work.}
\affiliation{Leibniz IFW Dresden, D-01069 Dresden, Germany}
\author{K.~Mehlawat}
\thanks{These authors contributed equally to this work.}
\affiliation{Leibniz IFW Dresden, D-01069 Dresden, Germany}
\affiliation{Institute for Solid State and Materials Physics and W{\"u}rzburg-Dresden Cluster of Excellence ct.qmat, TU Dresden, D-01062 Dresden, Germany}
\author{A. G. Moghaddam}
\affiliation{Leibniz IFW Dresden, D-01069 Dresden, Germany}
\affiliation{Department of Physics, Institute for Advanced Studies in Basic Sciences (IASBS), Zanjan 45137-66731, Iran}
\author{R. Ray}
\affiliation{Leibniz IFW Dresden, D-01069 Dresden, Germany}
\author{A. Zeugner}
\affiliation{Faculty of Chemistry and Food Chemistry, TU Dresden, D-01062 Dresden, Germany.}
\author{M. Richter}
\affiliation{Leibniz IFW Dresden, D-01069 Dresden, Germany}
\affiliation{Dresden Center for Computational Materials Science (DCMS), TU Dresden, 01062 Dresden, Germany}
\author{J. van den Brink}
\affiliation{Leibniz IFW Dresden, D-01069 Dresden, Germany}
\affiliation{Institute for Solid State and Materials Physics and W{\"u}rzburg-Dresden Cluster of Excellence ct.qmat, TU Dresden, D-01062 Dresden, Germany}
\author{A. Isaeva}
\affiliation{Leibniz IFW Dresden, D-01069 Dresden, Germany}
\affiliation{Institute for Solid State and Materials Physics and W{\"u}rzburg-Dresden Cluster of Excellence ct.qmat, TU Dresden, D-01062 Dresden, Germany}
\author{B.~B\"uchner}
\affiliation{Leibniz IFW Dresden, D-01069 Dresden, Germany}
\affiliation{Institute for Solid State and Materials Physics and W{\"u}rzburg-Dresden Cluster of Excellence ct.qmat, TU Dresden, D-01062 Dresden, Germany}
\author{V.~Kataev}
\affiliation{Leibniz IFW Dresden, D-01069 Dresden, Germany}

\begin{abstract}

\corr{The recent discovery of magnetic topological insulators has opened new avenues to explore exotic states of matter that can emerge from the interplay between topological electronic states and magnetic degrees of freedom, be it ordered or strongly fluctuating. Motivated by the effects that the dynamics of the magnetic moments can have on the topological surface states, we investigate the magnetic fluctuations across the \MBT\ family. 
Our paramagnetic electron spin resonance experiments reveal contrasting Mn spin dynamics in different compounds, which manifests in a strongly anisotropic Mn spin relaxation in \MnOTF\ while being almost isotropic in \MnOFS.  
Our density-functional calculations explain these striking observations in terms of the sensitivity of the local electronic structure to the Mn spin-orientation, and indicate that the anisotropy of the magnetic fluctuations can be controlled by the carrier density, which may directly affect the electronic topological surface states.} 
\end{abstract}

\date{\today}

\maketitle

\textit{Introduction.}
\corr{The experimental discovery of antiferromagnetic topological insulators (AFTIs) in the van der Waals \MBT\ family \cite{Otrokov2019,gong2019experimental} has provided a fertile new basis for the investigation of exotic phenomena rooted in the interplay between topology of the electronic structure and spontaneous symmetry breaking, such as the quantum anomalous Hall effect, the topological magnetoelectric effect and chiral Majorana fermions \cite{Chang2013,He2017,Tokura2019,jiang2020concurrence,deng2020quantum}.}
The two compounds studied the most so far, \MnOTF\ and \MnOFS, order antiferromagnetically with the A-type spin structure at N\'eel temperatures $T_{\rm N}\sim 24$\,K and $T_{\rm N}\sim 13$\,K, respectively \cite{chen2019intrinsic,souchay2019layered,PhysRevMaterials.3.064202,Ding2020,Wueaax9989,Vidal2019,hu2020van}. 
In this phase, both the time-reversal ($\Theta$) and a primitive-lattice translational ($T_{1/2}$) symmetries are broken but their combination $S=\Theta T_{1/2}$ is preserved \cite{PhysRevB.81.245209}.
\corr{For temperatures $T$ below $T_{\rm N}$, the topological protection of the surface states depends on whether or not the surface is $S$-symmetric.
At higher $T$,  even if $S$ can be broken locally and temporarily, the
surface spectrum is expected to be gapless for any surface due to the
restoration of $\Theta$ in a statistical sense
\cite{PhysRevB.89.155424,2020arXiv200303439G},
$\Theta$ being an average symmetry. As it results from averaging over magnetic fluctuations, the interplay between these and the topological electronic surface states naturally depends on space and time scales.}

\corr{Experimentally, surface states with an origin in the band inversion have been widely observed \cite{Otrokov2019,gong2019experimental,PhysRevB.100.121104,PhysRevX.9.041038,PhysRevX.9.041039,PhysRevX.9.041040,Wueaax9989,Vidal2019,hu2020van,PhysRevB.101.161109,PhysRevB.101.161113,gordon2019strongly,Klimovskikh2020,xu2019persistent,PhysRevX.10.031013,vidal2020orbital} but the details around the Dirac point are subject of controversy. In particular, among experiments that show a gapped surface spectrum for $T<T_{\rm N}$, whether the gap persists above $T_N$ represents an important open question. Our previous experimental results suggested that the dynamics of the bulk localized Mn magnetic moments play a key role in \MnOTF\ \cite{Otrokov2019} as electron spin resonance (ESR) measurements at $T > T_{\rm N}$ show that the relaxation of the Mn moments due to the exchange coupling with the conduction electrons is strongly anisotropic. This suggests a strong anisotropy of Mn spin fluctuations which may give rise to an instantaneous (on the timescale of ESR) polarization field  at the surface, preventing the gap to close even at $T \gg T_{\rm N}$ as observed on the much faster timescale of the ARPES experiment. Indeed further recent work proposes that the topological gap above $T_{\rm N}$ remains open due to short-range magnetic fields generated by chiral spin fluctuations \cite{Shikin2020}.}

\corr{Certainly, firm and final establishment of a correspondence between the characteristics of the magnetic fluctuations of both, conduction electrons and paramagnetic Mn spins, and the surface gap is a formidable task given the current, rather unsettled, experimental situation with regard to the surface band structure in different ARPES experiments triggering different, sometimes controversial,  interpretations. In this context it is of a paramount importance to unravel in detail the nature of the magnetic fluctuations across the \MBT\ family, the  understanding of which is currently lacking.} In this paper, we address this particularly crucial issue. First, we present ESR results which show that the anisotropy of the Mn spin relaxation rate in the paramagnetic state varies enormously in the family being unprecedentedly large in \MnOTF\ and nearly vanishing in \MnOFS. Second, based on density-functional calculations, we identify a critical role for the Mn spin dynamics played  by the magnetic anisotropy of the electronic structure, which consistently explains the ESR data and further suggests that in doped semiconductors the carrier density can be used to tune the high-temperature anisotropy of the spin dynamics.

\begin{figure*}[t]
	\includegraphics[width=18cm]{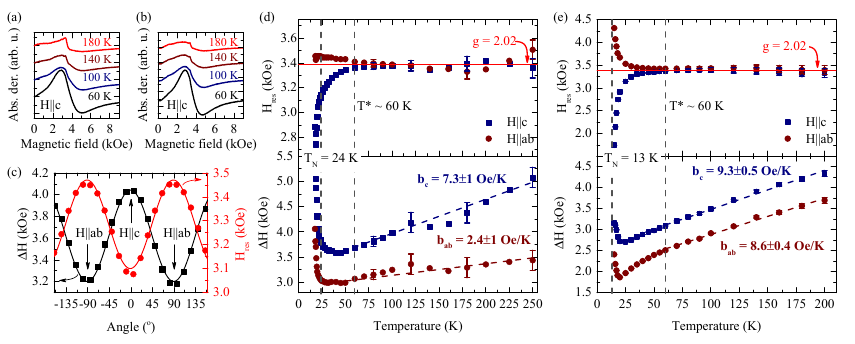}
	\caption{ESR measurements. (a),(b) Temperature dependence of the spectra (field derivatives of the microwave absorption) with \HIIc, for \MnOTF\ and \MnOFS, respectively.  (c) Angular dependence of the linewidth $\Delta H$ and of the resonance field $H_{\rm res}$ at $T=20$\,K for \MnOTF. Solid lines are fits of $\Delta H \sim [1+\cos^2(\theta)]$ and $H_{\rm res} \sim [1+\sin^2(\theta)]$, where $\theta$ is the angle between $H$ and the $c$~axis; (d),(e) Temperature dependence of  $H_{\rm res}$ and  $\Delta H$ measured with two fields orientations for  \MnOTF\ and \MnOFS, respectively.  Dashed lines are fits of the Korringa dependence $\Delta H(T) = \Delta H_0 + bT$ while vertical dashed lines indicate the N\'{e}el temperature $T_{\rm N}$ and a crossover characteristic temperature $T^\ast$ between two different types of temperature behavior of $H_{\rm res}$ and $\Delta H$.}
	\label{Fig_MnBi2Te4_1}
\end{figure*}

\textit{ESR experimental results.}
Electron spin resonance (ESR) measurements on high-quality \MnOTF\ and \MnOFS\ single crystalline samples synthesized and thoroughly characterized by X-ray diffraction and electron-dispersive X-ray spectroscopy in Refs.~\cite{Zeugner2019} and \cite{Vidal2019}, respectively,  were carried out at a microwave frequency of $\nu = 9.56$\,GHz and at temperatures from 4 to 300\,K using a commercial Bruker X-band spectrometer. The magnetic field $H$ was swept from 0 to 9\,kOe. The ESR signals have a well-defined asymmetric Dysonian shape \cite{Feher1955,Dyson1955}, typical for \corr{conducting} samples. \corr{This is indeed expected since Mn/Bi antisite intermixing is omnipresent in \MnOTF\ (Mn$_{0.85}$Bi$_{2.1}$Te$_4$) and \MnOFS\ (Mn$_{0.8}$Bi$_{4.1}$Te$_7$) crystals and acts as intrinsic self-doping \cite{Otrokov2019,Zeugner2019,Vidal2019,PhysRevLett.124.197201,PhysRevMaterials.3.064202,Ding2020}.} Characteristic spectra are presented in Fig.\ref{Fig_MnBi2Te4_1}(a),(b). At high temperature, the line shape gets somewhat distorted due to an emerging small and narrow impurity peak which can be easily taken into account in the Dysonian fit. In order to precisely align the samples in the magnetic field parallel (\HIIc) and normal (\HIIab) to the $c$~axis, we have measured the angular dependence of the linewidth ($\Delta H$) and of the resonance field ($H_{\rm res}$) at a low temperature, which are exemplarily plotted in Fig.~\ref{Fig_MnBi2Te4_1}(c). They follow a typical $\Delta H$$\sim$$[1$$+$$cos^2(\theta)]$ (or $H_{\rm res}$$\sim$$[1$$+$$sin^2(\theta)]$) dependence \corr{whose} extrema correspond to the respective field geometries, as indicated in the Figure. 

The $T$-dependence of $H_{\rm res}$ and of $\Delta H$ obtained from the Dysonian fits are shown for both field orientations on Fig.~\ref{Fig_MnBi2Te4_1}(d) for \MnOTF\ and on Fig.~\ref{Fig_MnBi2Te4_1}(e) for \MnOFS. Due to a decrease of the intensity of the ESR signal with increasing temperature according to the Curie-Weiss law, the error bars of the fit increase correspondingly.

One can identify in Figs.~\ref{Fig_MnBi2Te4_1}(d),(e) a crossover temperature $T^\ast \sim 60$\,K which separates two different types of behavior of $H_{\rm res}(T)$ and $\Delta H(T)$. \corr{At $T >T^\ast$, the resonance field is  $T$-independent and it has the same value (within error bars) for both directions of $H$}.  \corr{The resonance condition $h \nu  = g\mu _{\rm B} H_{\rm res}$ yields the isotropic $g$~factor $g = 2.02$ for both compounds.} It is close to the spin-only  value of \corr{$g_S \approx 2.0023$}, as expected for the Mn$^{2+}$ ($S$$=$$5/2$, $L$$=$$0$) ion. In contrast, at $T  < T^\ast$, $H_{\rm res}$ becomes significantly anisotropic. For \HIIab, the ESR line of \MnOTF\ shifts slightly to higher fields, whereas for \HIIc\ it shifts strongly to smaller fields. Such behavior is typical for the antiferromagnetic (AFM) resonance modes \cite{Turov}. However, these shifts commence well above  $T_{\rm N}\sim 24$ K and thus evidence the growth of the static, on the ESR time scale, short range Mn-Mn spin correlations already in the paramagnetic regime, which is typical for the intrinsically low-dimensional van der Waals magnets (see, e.g., Refs.~\cite{Zeisner2019,Zeisner2020}). At $T < T_{\rm N}$, the resonance line broadens and shifts further and finally disappears at $T \sim 17$\,K since  the AFM energy gap \cite{Turov} becomes larger than the microwave excitation energy $h\nu$ at $\nu = 9.56$\,GHz. \MnOFS\ exhibits at $T < T^\ast$ a qualitatively similar behavior [Fig.~\ref{Fig_MnBi2Te4_1}(e)]. In this case, the shifts are strong for both field geometries, typical for a ferromagnet with an easy axis anisotropy. Such ferromagnetic character of the low-$T$ ESR response of \MnOFS\ was established in our previous multifrequency ESR study \cite{Vidal2019}.

The central observation in our ESR experiments is the contrasting behavior of the resonance line parameters in the high-temperature regime above $T^\ast \sim 60$\,K. Here, an isotropic and $T$-independent value of the resonance field evidences an uncorrelated, paramagnetic state of the Mn spin system in both compounds. \corr{The linewidth follows a linear temperature dependence $\Delta H(T) = \Delta H_0 + bT$  [Figs.~\ref{Fig_MnBi2Te4_1}(d),(e)], characteristic of the Korringa relaxation of the localized magnetic moments by conduction electrons in the isothermal regime \cite{Barnes1981}.}
In this equation, $\Delta H_0$ is the $T$-independent residual width due to, e.g., \corr{spin-spin} interactions and magnetic field inhomogeneities, and \corr{the second term is the relaxation-driven $T$-dependent contribution 
\footnote{\corr{The linear in temperature Korringa relaxation is the unique fingerprint of the relaxation of the localized $d$-states via the exchange coupling to the conduction electrons. Other relaxation channels are either $T$-independent (e.g dipole-dipole or exchange interactions) or strongly non-linear in $T$ (spin-phonon relaxation) \cite{AbragamBleaney}. The latter mechanism is inactive for Mn$^{2+}$ due to the absence of the orbital moment.}}.}
Remarkably, for \MnOTF, \corr{the Korringa slope} $b = d \Delta H(T) / d T$  \cite{Korringa1950,Barnes1981} depends drastically on the direction of $H$. We obtain $b_{\rm c} = 7.3 \pm 1$\,Oe/K and $b_{\rm ab} = 2.4 \pm 1$\,Oe/K for \HIIc\ and \HIIab, respectively, yielding the ratio $b_{\rm ab} / b_{\rm c} = 0.33 \pm 0.14 \simeq 1/3$. In contrast, for \MnOFS, the values $b_{\rm c} = 9.3 \pm 0.5$\,Oe/K and $b_{\rm ab} = 8.6 \pm 0.5$\,Oe/K are very close, with the ratio $b_{\rm ab} / b_{\rm c} = 0.92 \pm 0.07$. 
While the overall larger $b$ values in \MnOFS\ are probably due to a higher doping level \footnote{This is supported by the difference in the room temperature resistivity values $\rho$ for these compounds: \MnOTF\ has $\rho \sim 1.5$ m$\Omega$cm \cite{Otrokov2019} and \MnOFS\ has $\rho \sim 0.31$ m$\Omega$cm \cite{Vidal2019}}, the very different anisotropy in the spin relaxation is more intriguing and will be discussed below.

\textit{Anisotropic Korringa relaxation. } In a magnetic resonance experiment, one measures the components of the dynamical magnetization $M_{x,y}$  of the paramagnetic species at their resonance frequency \cite{AbragamBleaney,Slichter96}, transversal to the applied static magnetic field $H$. According to the so-called Bloch-Wangness-Redfield (BWR) theory,  the width of the signal $\Delta H(T)$ is inversely proportional to the decay time $\tau$ of $M_{x,y}$  due to the relaxation processes caused by transversal and longitudinal fluctuating fields acting on the resonating spin ensemble \cite{Wangsness53,Bloch56,Redfield57,Slichter96}. In a metal, such fields are generated by conduction electrons exchange-coupled to the localized spins ($s$-$d$ coupling). This mechanism, known as the Korringa relaxation \cite{Korringa1950,Barnes1981}, gives rise to a linear in $T$ increase of the linewidth and reads, in the simplest case, $\Delta H(T) \sim 1/ \tau (T) \sim [J D(\varepsilon_F)]^2 T$ \cite{Note1}.
Here, $D(\varepsilon_F)$ is the density of states (DOS) of itinerant electrons at the Fermi energy $\varepsilon_F$ and $J$ is the $s$-$d$ coupling strength. In ordinary metallic systems, the Korringa slope $b = d \Delta H(T) / d T$ is found to be rather independent of the direction of $H$. There are only a few known examples of anisotropic Korringa relaxation in systems containing heavy elements with strong spin-orbit coupling (SOC) \cite{Vaknin1987,Vithayathil1991,Kataev2009}. These include OsF$_6$-graphite intercalated compound, where the anisotropy was attributed to the anisotropy of $J$ \cite{Vaknin1987}, and UPt$_3$, where an anisotropic relaxation rate in a nuclear magnetic resonance experiment \cite{Vithayathil1991} was explained by the anisotropic fluctuations of the heavy-fermion spins. Notably, in both cases one finds the ratio of the Korringa slopes $b_{\rm ab}/b_{\rm c} \gtrsim 1/2$ \footnote{Note, that for the OsF$_6$-GIC this ratio $b_{\rm ab}/b_{\rm c} \gtrsim 1/2$ is still true if the g-factor anisotropy, present in this compound, is not taken into account.}. In this respect, the significantly smaller ratio $b_{\rm ab} / b_{\rm c} \simeq 1/3$ found in \MnOTF\ is unprecedented and remarkable.

To understand the sources of anisotropy in the Korringa relaxation, it is convenient to express the relaxation rate in terms of the frequency $\omega$- and wave vector ${\bf q}$-dependent spin susceptibility $\chi_{i,i}(\omega,{\bf q})$ of conduction electrons, with $i=x,y,z$. Choosing the $z$ axis along the direction of $H$, the relaxation rate reads \cite{moriya63,narath72}
\begin{equation}
\frac{1}{\tau_z}= T\,
\lim_{\omega\to0}\:
\frac{1}{\omega}
\sum_{\bf q} {\rm Im}\: 
\big[
2\, J_{z}^2 \,
\chi_{z z}(\omega,{\bf q}) 
+
\sum_{i=x,y}
J_{i}^2 \,
\chi_{ii}(\omega,{\bf q}) 
\big],
\end{equation}
where we have set $\hbar=k_{\rm B}=1$ and we have considered that the $s$-$d$ exchange coupling consists of three different diagonal components $J_i$. An explicit calculation, see Supplementary Materials (SM) \footnote{See Supplementary Materials for: $(i)$ derivation of Eq. \ref{perp-final}, $(ii)$ a discussion on the shape factors and $(iii)$ further details on the DFT results.}, yields
\begin{eqnarray}
\label{perp-final}
&&\frac{1}{\tau_z}= \pi T 
\big[
2\,J_{z}^2 \: |{\cal S}_{z}(\varepsilon_F)|^2
+
\sum_{i=x,y} 
J_{i}^2 \: |{\cal S}_{i}(\varepsilon_F)|^2
\big]
,\\
&&
\label{perp-final2}
|{\cal S}_{i}(\varepsilon)|^2=
\sum_{\nu{\bf k}, \nu'{\bf k}'} 
\delta(\varepsilon - \varepsilon_{\nu{\bf k}}) 
\delta(\varepsilon - \varepsilon_{\nu'{\bf k}'})
|{\cal F}^{i,\nu\nu'}_{{\bf k}{\bf k}'}|^2,
\end{eqnarray}
where $\nu$ is a band index, {\bf k} the momentum and  
$|{\cal F}_{{\bf k}{\bf k}'}^{i,\nu\nu'}|^2$  denotes 
${\rm Tr}\big[ {\cal F}_{{\bf k}{\bf k}'}^{i,\nu\nu'} {\cal F}_{{\bf k}'{\bf k}}^{i,\nu'\nu} \big]$,
with 
${\cal F}^{i,\nu\nu'}_{{\bf k}{\bf k}'}=  \langle u_{\nu{\bf k}} |\hat{\sigma}_i |u_{\nu'{\bf k}'} \rangle/2 $ the matrix elements of the spin operator $ \hat{\sigma}^i $ with respect to the spinor parts of Bloch functions $|u_{\nu{\bf k}}\rangle$$=$$a_{\nu{\bf k}}|$$\uparrow\rangle+b_{\nu{\bf k}}|$$\downarrow\rangle$. 
Without SOC, ${\cal F}^{i,\nu\nu'}_{{\bf k}{\bf k}'}$  is momenta-independent 
and, therefore, ${\cal S}_i(\varepsilon)\equiv D(\varepsilon)$. If we further consider an isotropic $J$, the original Korringa result is recovered \cite{Korringa1950}. In general, SOC can give rise to anisotropy both in $J_i$ and in ${\cal S}_i(\varepsilon_F)$, which in turn may lead to anisotropic relaxation rates, as indicated by Eq. (\ref{perp-final}).

Eq.~(\ref{perp-final}) was obtained in the zero magnetic field limit, which implies that the Larmor frequency of Mn spins $\omega_{\rm L}$ is much smaller than the frequency of the spin fluctuations of the conduction electrons $\omega_{\rm fl}$. In this limit, Eq.~(\ref{perp-final}) matches the BWR results expressing the fluctuating fields introduced in this theory as $\tau_0\overline{H^2_i}  = T  J_{i}^2 \: |{\cal S}_{i}(\varepsilon_F)|^2$ (see SM \cite{Note4}), where the correlation time $\tau_0$ in our problem is $1/\omega_{\rm fl}$. This connection provides a physical interpretation for the shape factors ${\cal S}_i$: They are the band structure property that together with $T$ and $J_i$ determine the effective magnetic field exerted by the electron cloud on the Mn spins on a timescale $\tau_0$. It further suggests an extension to the finite field case, $\omega_{\rm L}\lesssim \omega_{\rm fl} $, which in the BWR theory yields a pre-factor ${\cal A}_\omega\sim (1+\omega_L^2/\omega_{\rm fl}^2)^{-1}$ for the second term in Eq.~(\ref{perp-final}) \cite{Slichter96}.

Given the layered structure of these compounds, $J_c$ and ${\cal S}_{c}$ can be significantly different from their in-plane counterparts $J_{ab}, {\cal S}_{ab}$ \footnote{We neglect anisotropy within the $ab$ plane in all discussions and calculations.}. In this case, the ratio of the relaxation times reads
\begin{equation}
\frac{\tau_c}{\tau_{ab}}=\frac{1}{2}\frac{2+{\cal A}_\omega(1+\zeta)}{{\cal A}_\omega+\zeta}\:,
\label{tau_ratio}
\end{equation}
where $\zeta=|J_{c}{\cal S}_{c}(\varepsilon_F)/J_{ab}{\cal S}_{ab}(\varepsilon_F)|^2$.
If ${\cal S}_{i}(\varepsilon_F)$ is isotropic and $\omega_L/\omega_{\rm fl}\ll1$, the remaining anisotropy due to that of $J_i$ is bounded as $(1/2)_{J_{ab}\rightarrow 0} <\tau_c/\tau_{ab}<(3/2)_{J_c\rightarrow 0}$, posing a problem to account for the experimental result for \MnOTF,  $b_{\rm ab} / b_{\rm c} = \tau_c/\tau_{ab} \simeq 1/3$. Assuming a finite ratio $\omega_L/\omega_{\rm fl}$ which yields ${\cal A}_{\omega}<1$, the above limits can be surpassed, ultimately reaching $\tau_c/\tau_{ab} = 1/4$ for $\omega_{\rm L} = \omega_{\rm fl}$ and $\zeta \gg1$. However, that would require $J$ to have an extreme Ising-like anisotropy with vanishing $J_{ab}$ in \MnOTF\ while being almost isotropic in \MnOFS.

\begin{figure}[t!]
\centering
\includegraphics[width=8.5cm,angle=0]{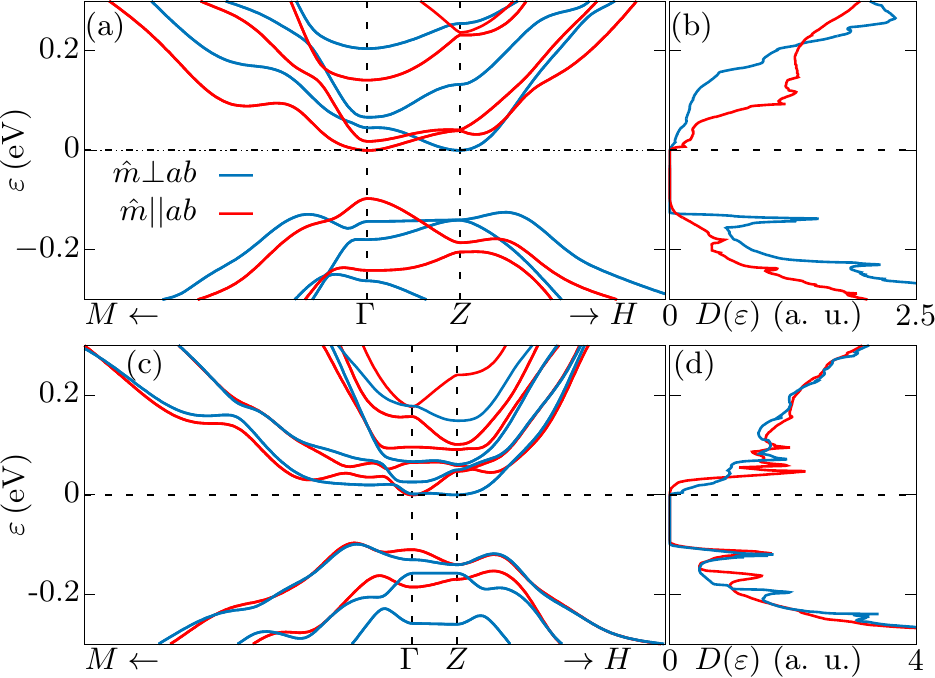}
	\caption{(a),(b) Band structure and density of states, respectively, of \mbt\ in the AFM phase for out-of-plane (blue) or in-plane (red) \corr{Mn} magnetic moment direction.
	(c),(d) Analogous results for \mbtl. }
\label{fig:dft} 
\end{figure}

Therefore, to understand the anisotropy of the Korringa relaxation in \MnOTF, it is necessary to analyze a possible anisotropy of the form factors ${\cal S}_i(\varepsilon_F)$ associated with the conduction electron cloud. This can be relevant when the carriers have a large SOC, as in the studied compounds. In such a case, for temperatures where the timescale of \corr{the dynamics of the Mn spins} is longer than the timescale of conduction electrons, one may expect changes in the local electronic structure arising from the induced instantaneous polarization of the local moments, "frozen" on the fast electronic timescale, which can in turn affect the relaxation.

\textit{Band structure anisotropy. }
We now explore this idea, assuming a large separation between timescales: 
\corr{The measured ESR occurs on the slowest scale ($\sim10^{-10}$ s);
on the intermediate scale ($ \sim 10^{-13}$\ldots $10^{-14}$ s), the Mn magnetic moment direction $\hat{m}$ fluctuates; on the fast scale ($\sim 10^{-15}$ s), the electronic system adapts to the local Mn spin direction and dissipates the excitation.} Figs.~\ref{fig:dft}(a),(b) present the energy bands and DOS of \MnOTF\ in the AFM phase for different $\hat{m}$, and Figs.~\ref{fig:dft}(c),(d) show results for \MnOFS\ \footnote{Fully relativistic calculations in the Generalized Gradient Approximation+U (atomic limit) were performed using  the FPLO code version 18.52 with parameters $U=2$\,eV and $J=1$\,eV. We used a linear tetrahedron method with k-meshes of $24\times24\times24$ (rhombohedral setup) and $22\times22\times4$ (hexagonal setup) subdivisions for \MnOTF\ and   \MnOFS, respectively. To compute the DOS, we used in each case meshes of up to $60\times60\times60$ subdivisions and  $60\times60\times5$. Structural details are based on experimental
lattice parameters, see Refs. [1] and [13]. }. The sensitivity to $\hat{m}$ is significantly smaller in \MnOFS. For this compound, within planes perpendicular to $\Gamma Z$, the bands are much less affected by $\hat{m}$, which naturally reflects in $D(\varepsilon)$. Such lesser sensitivity possibly originates in various "dilution" effects caused by the Mn-free quintuple layer in \MnOFS\ \cite{Note4}. 

\begin{figure}[t!]
\centering
\includegraphics[width=8.5cm,angle=0]{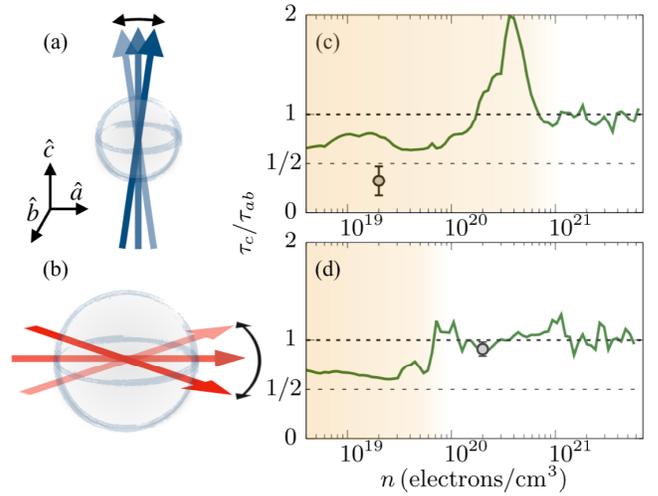}
	\caption{\corr{(a,b) Pictorial representation of the anisotropic spin dynamics. An out-of-plane Mn spin is accompanied by a relatively small carriers DOS (gray electron cloud) and, accordingly, has a relatively slow dynamics as compared to an in-plane Mn spin, accompanied by a larger DOS and presenting faster dynamics.}
(c,d) Estimation of relaxation rate ratio as a function of the carrier density for \mbt (c) and for \mbtl (d). The gray circles correspond to the experimental ESR data, estimating $n$ from the Hall measurements on the same samples in Refs.~\cite{Otrokov2019} and \cite{Vidal2019}, respectively. 
	}
\label{fig:dft2} 
\end{figure}

In \MnOTF, for small electron doping, the DOS tends to be larger when $\hat{m}||ab$. This tendency qualitatively matches the results of the ESR experiment, where, for $H||c$, \corr{the in-plane part of the Mn dynamic magnetization $M_{x,y} = M_{a,b}$ is probed}. For those Mn ions that have their spin essentially in-plane, the local electronic structure near them is better described by the calculation with $\hat{m}||ab$. The larger DOS in such case contributes to a faster relaxation rate \corr{of $M_{x,y}$} than in the configuration $H||ab$\corr{, where $M_{x,y} = M_{a,c} = M_{b,c}$ is probed} [Figs.~\ref{fig:dft2}(a),(b)].

These ideas can be illustrated with the following ansatz for the shape factors \cite{Note4}.
We estimate them as ${\cal S}_i$$=$$\langle D^2_{ab} \rangle$ for $H||c$ and as ${\cal S}_i$$=$$\langle D^2_{bc} \rangle$ for $H||ab$, where $\langle D^2_{ij} \rangle $ is the squared effective DOS when $\hat{m}$ is along the direction $i$ or $j$.
Since the shape factors are defined by the electronic structure, it is worth stressing that the dependence on the direction of $H$ does not come from the negligible $H$-dependence of the band structure. 
Rather, the role of $H$ -- together with the radio-frequency field direction -- \corr{is to \textit{select} the particular kind of fluctuations}. For different directions of $H$, the measured resonance originates in Mn ions with different $\hat{m}$. The associated shape factors, considering the locally distinct electronic structure, are accordingly different.

Neglecting a possible anisotropy of $J$ and assuming the limit $\omega_L/\omega_{\rm fl}\ll1$, so that the effects described are purely related to the band structure, this approach yields $\tau_c/\tau_{ab} = \langle D^2_{bc} \rangle / \langle D^2_{ab}\rangle$.
Figs.~\ref{fig:dft2}(c),(d) show the estimated ratio $\tau_c/\tau_{ab}$ as a function of the carrier density $n$ for \MnOTF\ and \MnOFS, respectively. Specifically, we consider $\langle D^2_{ab} \rangle$$=$$D^2_{\hat{m}||ab}$ and $\langle D^2_{bc} \rangle$$=$$(D^2_{\hat{m}||ab}$$+$$D^2_{\hat{m}||\hat{c}})/2$, where the DOS are evaluated at fixed $n$. 
For comparison, our experimental ESR results are included as gray circles.  The almost isotropic Korringa relaxation in \MnOFS\ is well captured, whereas the anisotropy in \MnOTF\ is underestimated. However, quantitative agreement may be achieved by considering an anisotropy of $J$ or a finite Larmor frequency, as discussed above.

Interestingly, both compounds exhibit a crossover as a function of doping signalled by large deviations from $\tau_c/\tau_{ab}$$=$$1$ at small $n$ to nearly vanishing anisotropy at large $n$. It appears that the main difference between the compounds is the level of doping at which the crossover occurs. In particular, the samples of \MnOTF\ and  \MnOFS\ lie on the anisotropic and isotropic sides of this crossover, respectively, in agreement with the ESR results. Reducing $n$ in \MnOFS\ by only one order of magnitude, potentially achievable by partial substitution of Bi by Sb \cite{chen2019intrinsic}, would recover the strong anisotropy $\tau_c$$<$$\tau_{ab}$, while a significant increase of $n$ in \MnOTF\ would change its type ($\tau_c$$>$$\tau_{ab}$) and eventually turn it to the isotropic limit ($\tau_c$$=$$\tau_{ab}$). 

\corr{These results should be relevant for the interpretation of experiments in which the observed physical processes are not slow enough to average out the anisotropy of the Mn spin dynamics. This is for instance the case for the photoemission process in the non-adiabatic, sudden regime, which is typically associated with a timescale faster than $10^{-15}$\,s. In particular, the much slower out-of-plane Mn spin dynamics in MnBi$_2$Te$_4$ indicates a larger out-of-plane dynamic polarization yielding a non zero instantaneous magnetic field perpendicular to the sample surface, which could contribute to the persistence of an electronic gap in the paramagnetic phase.
Our theoretical results suggest that appropriate engineering of the carrier density can favor the opposite or vanishing anisotropy and hence, may provide a way to experimentally control the surface topological spectrum.}

\textit{Conclusions. }
\corr{Our combined ESR and theoretical study shows that the Mn spin dynamics in magnetic topological insulators can be strongly anisotropic. The large spin-orbit coupling of conduction electrons plays a key role, making the local electronic structure strongly sensitive to Mn spin rotations, which in turn affects the Mn spin relaxation. Since the Mn spin dynamics can affect the topological surface states, our finding of the carrier density as a knob to control the anisotropy of magnetic fluctuations suggests a new way to tune the high-temperature surface spectrum of magnetic topological insulators. Altogether our results open a new perspective for exploring the magnetic dynamics and its interplay with non-trivial electronic structure in magnetic topological insulators and call for further investigations of the topological surface states at different time scales.}

\textit{Acknowledgments. }
This work was financially supported by the Deutsche Forschungsgemeinschaft (DFG) through grant No. KA1694/12-1 and within the Collaborative Research Center SFB 1143 ``Correlated Magnetism – From Frustration to Topology” (project-id 247310070) and the Dresden-Würzburg Cluster of Excellence (EXC 2147) ``ct.qmat - Complexity and Topology in Quantum Matter" (project-id 39085490). J.I.F. acknowledges the support from the Alexander von Humboldt Foundation. K. M. \mbox{acknowledges} the Hallwachs–Röntgen Postdoc Program of ct.qmat for financial support. We thank Ulrike Nitzsche for technical assistance.

\bibliography{mybibfile}

\clearpage

\pagebreak

\onecolumngrid
\section{Supplementary Material}

\subsection{ Details on the theory of ESR for spin-orbit coupled systems}

Following the well-established theory of NMR, when the static field is applied along $z$-axis, the spin relaxation rate can be expressed as \cite{moriya63,narath72}
\begin{eqnarray}
\frac{1}{\tau_z}
&=&
\frac{\gamma^2}{4}\int dt \: \cos\omega_L t \langle \{\delta B_+(t) ,\delta B_-(0) \}  \rangle +
\frac{\gamma^2}{2}\int dt \, \langle \{\delta B_z(t) ,\delta B_z(0) \}  \rangle~,
\label{tau}
\end{eqnarray}
in terms of the fluctuating internal field $\delta {\bf B}$ acting on the local moment. 
Here $\gamma$ denotes the gyromagnetic ratio, $\omega_L=\gamma H_{\rm res}$ is the resonance frequency,  ${\hat{A},\hat{B}}=\hat{A}\hat{B}+\hat{B}\hat{A}$ is the anticommutator of two operators, and finally $\langle \hat{A} \rangle = {\rm Tr}(\rho_{\rm th} \hat{A} )$ denotes the statistical average of the operator. Also the field is decomposed as $\delta {\bf B}=\delta B_+ \hat{\bf e}_- + \delta B_- \hat{\bf e}_+ + \delta B_z \hat{\bf e}_z$ with $\delta B_\pm = \delta B_x \pm i \delta B_y$ and 
$\hat{\bf e}_\pm = (\hat{\bf e}_x \pm i \hat{\bf e}_y)/2$. Notice that $\hat{\bf e}_\pm$ and $\hat{\bf e}_z$ also create an orthogonal coordinate frame.

We consider a bath of noninteracting conduction fermions ${\cal H}_0 = \sum_{\nu k} \varepsilon_{\nu {\bf k} } c^\dag_{\nu {\bf k}} c_{\nu {\bf k}}$,
where $\varepsilon_{\nu {\bf k}}$ are the energy bands of the corresponding Bloch wavefunctions $| \psi_{\nu {\bf k}} \rangle = e^{i{\bf k}\cdot {\bf r}}| u_{\nu {\bf k}} \rangle $. 
These fermions are coupled  with a local moment ${\bf S}$ via a general anisotropic $s$-$d$ exchange coupling, which reads
\begin{eqnarray}
{\cal H}_{\rm ex}= \sum _{\bf q} {\bf S}\cdot {\bf J} \cdot \hat{\bm s}_{\bf q}
=\sum _{ij,{\bf q}}  J_{ij}  {S}^i  \hat{s}_{\bf q}^j, \text{ with } \hat{s}^j_{\bf q}=\frac{1}{2}\sum_{{\bf k},s,s'} c^\dag_{{\bf k}+ {\bf q},s} \sigma^j_{ss'} c_{{\bf k},s'}
\end{eqnarray}
Here, we omit, for brevity, the band index.
The fluctuating fields due to the electronic cloud around the local moments 
and in terms of electron creation and annihilation field operators,
read
\begin{eqnarray}
\widehat{\delta B}_\pm &=& \frac{1}{\gamma}\sum_{j,{\bf q}} J_{\pm,j}  \hat{s}_{\bf q}^j = \frac{1}{\gamma}\sum_{j,{\bf q}} 
\big( J_{x,j} \pm i J_{y,j} \big)  \: \hat{s}_{\bf q}^j \label{fluc-field}\\
\widehat{\delta B}_z &=& \frac{1}{\gamma} \sum_{j,{\bf q}} J_{z,j} \: \hat{s}_{\bf q}^j
\label{fluc-field2}
\end{eqnarray}
Inserting above expressions in Eq. \ref{tau}, we find:
\begin{eqnarray}
\frac{1}{\tau_z}
&=&
\frac{1}{4} \sum_{i,j}
J_{+,i} J_{-,j} 
\int dt \:\cos\omega_L t 
\sum_{\bf q}
\langle \{ \hat{s}_{\bf q}^i(t), \hat{s}_{\bf -q}^{j} \}  \rangle 
+ 
\frac{1}{2} \sum_{i,j}
J_{z,i} J_{z,j} 
\int dt
\sum_{\bf q}
\langle \{ \hat{s}_{\bf q}^i(t), \hat{s}_{\bf -q}^{j} \}  \rangle 
\end{eqnarray}
Using the symmetry of the anticommutator, $\langle \{ \hat{s}_{\bf q}^i(t), \hat{s}_{\bf -q}^{j} \}  \rangle =
\langle \{ \hat{s}_{\bf -q}^j(-t), \hat{s}_{\bf q}^{i} \}  \rangle$,
and noting that   $\frac{J_{+,i}\: J_{-,j}+J_{-,i}\: J_{+,j}}{2}
=
J_{x,i}\:  J_{x,j} + J_{y,i}\: J_{y,j}$
the relaxation time reads
\begin{eqnarray}
\frac{1}{\tau_z}&=&
T\:\sum_{i,j}
\big(J_{x,i}\:  J_{x,j} + J_{y,i}\: J_{y,j}+2 \, J_{z,i}\: J_{z,j} \big)
\times \lim_{\omega\to0}\:
\frac{1}{\omega}
\sum_{\bf q} {\rm Im}\: \chi_{ij}(iq_n,{\bf q})|_{iq_n\to \omega+i0^+}
\end{eqnarray}
To obtain this equation, we have used the quantum-mechanical version of fluctuation-dissipation theorem
defining the spin susceptibility tensor ${\bm \chi}(iq_n,{\bf q})$ with components $\chi_{ij}(iq_n,{\bf q})$. This relation is valid for large enough temperature $T\gg\omega_L$. Note that we have set the Planck and Boltzmann constant as $\hbar=k_B=1$.
The spin susceptibility in Matsubara representation is
\begin{eqnarray}
\chi_{ij}(iq_n,{\bf q})=-\frac{T}{4}
\sum_{i k_n , {\bf k}}{\rm Tr} \big[
\hat{\sigma}^i \:\hat{G}(ik_n,{\bf k}) \:\hat{\sigma}^j \:\hat{G}(ik_n+iq_n,{\bf k}+{\bf q}) 
\big],
\label{matsubara-chi}
\end{eqnarray}
with bosonic and fermionic Matsubara frequencies denoted as $q_n=2n\pi T$ and $k_n=(2n+1) \pi T$, respectively.
The noninteracting Green's function can be generally written as
\begin{eqnarray}
\hat{G}(ik_n,{\bf k}) = \sum_{\nu}\frac{| u_{\nu {\bf k} } \rangle \langle u_{\nu {\bf k}} | }{ik_n-\varepsilon_{\nu{\bf k}}}.
\label{matsubara-green}
\end{eqnarray}
We can readily arrive at the following suitable expression for the susceptibility
\begin{eqnarray}
\chi_{ij}(iq_n,{\bf q})=-\frac{T}{4}\sum_{i k_n , {\bf k}} \sum_{\nu\nu'}{\rm Tr} \big[
{\cal F}^{i,\nu\nu'}_{{\bf k},{\bf k}+{\bf q}}\: {\cal F}^{j,\nu'\nu}_{{\bf k}+{\bf q},{\bf k}}\big]
\times\frac{1
}
{(ik_n-\varepsilon_{\nu{\bf k}})(ik_n+iq_n-\varepsilon_{\nu',{\bf k}+{\bf q}})},
\label{chi-overlap}
\end{eqnarray}
with ${\cal F}^{i,\nu\nu'}_{{\bf k}\,{\bf k}'}=\langle u_{\nu{\bf k}} |\hat{\sigma}^i | u_{\nu'{\bf k}' } \rangle/2$.
Then by performing the Matsubara sum we find 
\begin{eqnarray}
	\frac{1}{\tau_z}= {-\pi T} \:\sum_{i,j}
	\big(J_{x,i}\:  J_{x,j} + J_{y,i}\: J_{y,j} +2 \, J_{z,i}\: J_{z,j}  \big)
	\sum_{{\bf k} , {\bf q} , \nu,\nu'} 
	\delta(\varepsilon_{\nu{\bf k}} -
	\varepsilon_{\nu',{\bf k}+{\bf q}}) \:
	\frac{\partial n(\omega)}{\partial \omega}\bigg|_{\omega=\varepsilon_{\nu{\bf k}}}
	{\rm Tr}\big[{\cal F}^{i,\nu\nu'}_{{\bf k},{\bf k}+{\bf q}}\: {\cal F}^{j,\nu'\nu}_{{\bf k}+{\bf q},{\bf k}}\big]
	.
\end{eqnarray}


Considering diagonal exchange coupling terms $J_{ij}=J_i \delta_{ij}$, defining ${\rm Tr}\big[ {\cal F}_{{\bf k}{\bf k}'}^{i,\nu\nu'} {\cal F}_{{\bf k}'{\bf k}}^{i,\nu'\nu} \big]=|{\cal F}_{{\bf k}{\bf k}'}^{i,\nu\nu'}|^2$ and replacing the derivative of Fermi-Dirac distribution $n(\omega)$ with its zero temperature value, 
this simplifies to
\begin{eqnarray}
\frac{1}{\tau_z} &=& \pi T \:
\sum_{\nu{\bf k} , \nu'{\bf k}' } 
\delta(\varepsilon_{\nu{\bf k}} -
\varepsilon_{\nu'{\bf k}'}) 
\:
\delta(\varepsilon_F-\varepsilon_{\nu{\bf k}}) 
\times \bigg(
J_x^2 |{\cal F}_{{\bf k}{\bf k}'}^{x,\nu\nu'}|^2
+
J_y^2 |{\cal F}_{{\bf k}{\bf k}'}^{y,\nu\nu'}|^2
+
2
J_z^2 |{\cal F}_{{\bf k}{\bf k}'}^{z,\nu\nu'}|^2
\bigg),~~
\label{perp}
\end{eqnarray}
which is the result presented in Eq. 2 and 3 of the main text.

\subsection{On the shape factors ${\cal S}_i$}

Here, first, for completeness, we recall the essential results from the Bloch-Wangness-Redfield (BWR) theory; and second, we further motivate the approximation proposed for the shape factors ${\cal S}_i$.
The BWR theory analyzes the relaxation phenomena in an ensemble of magnetic moments subject to a random magnetic field $H_j(t)$, with $j=x,y,z$  \cite{Slichter96}. 
Considering the external static field $H$ along $\hat{z}$ and the external radiofrequency field $H_{\rm rf}$ within the $xy$ plane, the relaxation time results
\begin{equation}
\frac{1}{\tau_z} = \gamma^2 \Big( \tau_0 \overline{H}^2_z  + \frac{1}{2}\frac{\tau_0}{1+\omega_{L}^2\tau_0^2} (\overline{H}^2_x+\overline{H}^2_y) \Big).
\label{BWR}
\end{equation}
Here, $\tau_0$ and $\overline{H}^2_j$ characterize the correlation in time of the random forces, $\overline{H_j(t) H_j(t+\tau)} = \overline{H_j}^2 {\rm exp}(-\tau/\tau_0)$.
The asymmetric appearance of the different field components in Eq. \ref{BWR} can be traced to their different role, which roughly speaking can be thought of as randomizing the Larmor frequency (the $z$-component) or the radiofrequency field (the $xy$ components).

In our case, the random magnetic fields are given by the exchange with conduction electrons [Eqs. \ref{fluc-field},\ref{fluc-field2}] and $\tau_0$ can be related with the frequency of the spin fluctuations of these electrons, $\tau_0 \sim 1/\omega_{fl}$. 
As shown in Section \textbf{(I)}, while these random magnetic fields naturally depend on the crystal momentum of the carriers, their net effect on the relaxation rate is encoded in integral properties of the band structure --the shape factors ${\cal S}_i$. 
A comparison of Eqs. 2 and 3 of the main text with Eq. \ref{BWR} suggests the identification
$\tau_0 \overline{H}_j^2 = T J_j^2 {\cal{S}}_j^2$,
which provides a simple physical interpretation: the effective magnetic field exerted by the conduction electronic cloud over the timescale $\tau_0$ is determined by the $s$-$d$ exchange, the temperature and the shape factors ${\cal S}_i$.

Our DFT results indicate that due to the spin-orbit coupling the electronic structure, e.g. the density of states, presents a strong sensitivity to the direction of the Mn magnetic moments $\hat{m}$. The ansatz for the shape factors proposed in the main text aims to describe how such sensitivity can in turn affect the local moment relaxation process. 
Notice that strictly speaking, such effects are absent in Eq. \ref{perp} since correlations between the conduction electrons and the magnetic moments were neglected.

While a complete theory is beyond the scope of this work, a heuristic motivation for the ansatz presented in the main text is based on the following considerations. First, one considers that due to the spin-orbit coupling the instantaneous magnetic field exerted by the conduction electron cloud on the Mn moments depends on $\hat{m}$. Defining an instantaneous shape factor $s_j(t,\hat{m})$, one has
\begin{equation}
T J_j^2 {\cal{S}}_j^2 =  T J_j^2 \int_0^{\tau_{ce}} dt\, s_j(t,\hat{m}),
\end{equation}
where $\tau_{ce}$ is the timescale associated with conducting electrons, typically $\sim \hbar/W$, with $W$ their bandwidth.
Second, the timescale of Mn spin fluctuations is assumed to be \textit{larger} than $\tau_{ce}$ and naturally smaller than $\tau_z$. In this intermediate timescale local rotations of $\hat{m}$ are followed essentially instantaneously by the electron cloud.
Third, out of the ensemble of fluctuating Mn spins, a given configuration of $H$ and $H_{\rm rf}$ picks up only certain components of the dynamical magnetization. For instance, for $H || c$ and $H_{\rm rf} || ab$, the measured relaxation originates in the in-plane component of the dynamical magnetization.
Thus, we assume the electron cloud near the contributing Mn moments -- and, in turn, the exerted magnetic field -- to be better described by the electronic structure with in-plane magnetic moments.

Last, an argument for the estimation of the shape factor as the DOS computed for different $\hat{m}$ is based on considering the effects of the SOC perturbatively. At zero order, ${\cal S}^2_i \propto D^2(\varepsilon_F)$. At first order, the matrix elements in $|{\cal F}_{{\bf k}{\bf k}'}^{i,\nu\nu'}|^2$ in Eq. \ref{perp} are unaffected and the only effect on ${\cal S}^2_i$ is caused by the change in the eigenenergies. This change reflects in the electronic DOS in a manner that depends on the magnetic moment orientation. Our ansatz, $S_i^2 \propto D^2(\varepsilon_F)|_{\hat{m} \perp H}$, can be thought of as using this first-order perturbation theory result replacing the DOS with the one computed with DFT for different $\hat{m}$.

\subsection{Details of the DFT results}

\subsubsection{On the different electronic structure anisotropy of \MnOTF\ and \MnOFS}

To further understand the origin of the different sensitivity of the density of states of MnBi$_2$Te$_4$ and MnBi$_4$Te$_7$ to the orientation of the magnetic moments, here we consider the projection of the density of states on different atoms in the unit cell. For brevity, we define $\Delta = D_{\mathbf{m}||ab} - D_{\mathbf{m}||c}$. 
The key structural difference between these compounds is the additional quintuple layer (QL) in \MnOFS. One could, therefore, suspect that the contribution to the Bloch states of the atoms in the QL could be less sensitive to $\hat{m}$ than the contribution of the atoms in the septuple layer (SL). The additional QL would, therefore, act to reduce (by diluting) the overal sensitivity of the total density of states. Fig. \ref{fig:dft3}(a) shows the projection of $\Delta$ on the QL (named $\Delta^{\text{QL}}$) and on the SL ($\Delta^{\text{SL}}$) and it can be seen that in the range of energies $[0.1,0.3]$\,eV, $|\Delta^{\text{QL}}| < |\Delta^{\text{SL}}|$. The difference are, however, rather modest. 
More significant than this is the effect that the additional QL has on the sensitivity of the SL contribution. To illustrate this, Fig. \ref{fig:dft3}(b) shows the projection of $\Delta$ on the Bi atoms in the SL, both for \MnOTF\, ($\Delta^{\text{Bi}}_{124}$) and \MnOFS\, ($\Delta^{\text{Bi}}_{147}$). 
It can be seen that $|\Delta^{\text{Bi}}_{124}|$ is significantly larger than $|\Delta^{\text{Bi}}_{147}|$. 
A similar reduction is found in the projection on the Te atoms in the SL (not shown).

\begin{figure}[h!]
	\centering
	\includegraphics[width=7.5cm,angle=0]{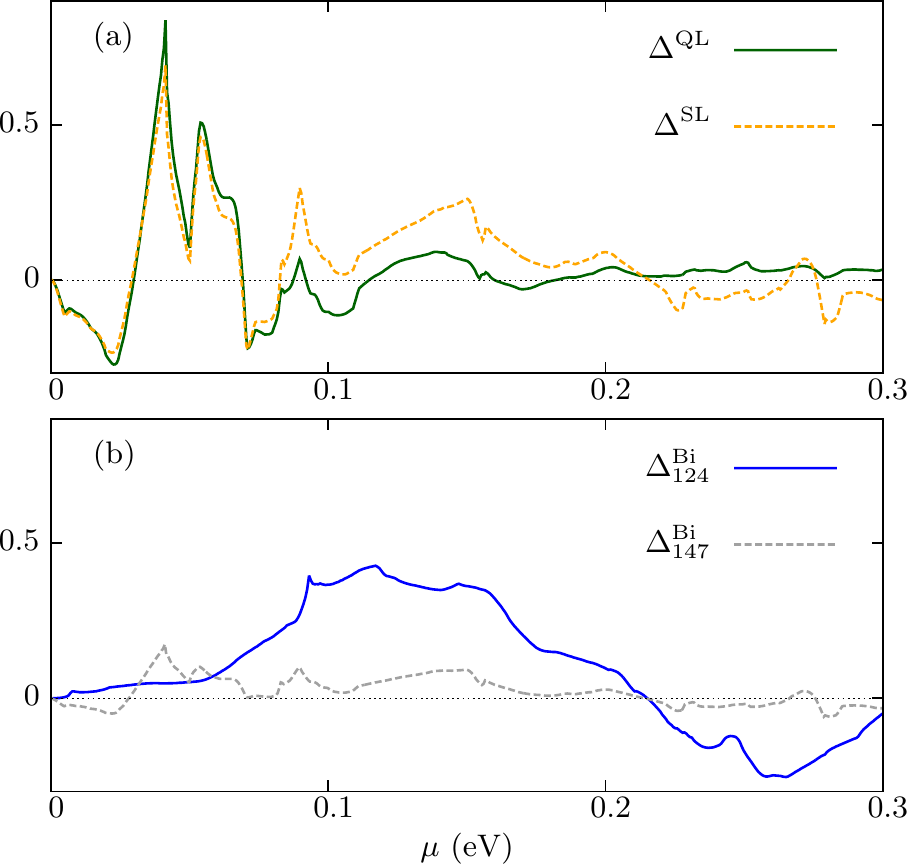}
	\caption{$\Delta$ is the difference between the density of states corresponding to the cases $\hat{m}||ab$ and $\hat{m}||c$. (a) For MnBi$_4$Te$_7$, projection of $\Delta$  on the quintuple and sextpule layers.
		(b)  For MnBi$_2$Te$_4$ and MnBi$_4$Te$_7$, projection of $\Delta$ on the Bi atoms in the septuple layer.
	}
	\label{fig:dft3} 
\end{figure}

\subsubsection{Carriers density vs Fermi energy.}

For the estimation of $\tau_c/\tau_{ab}$ as a function of the carrier density $n$, we consider the involved DOS at fixed $n$. For this, we first compute $n$ as a function of the Fermi energy $\varepsilon_F$. These curves, shown in Fig. \ref{fig:dft3} also make very transparent the sensitivity of the electronic structure to the direction of the magnetic moments $\hat{m}$. 
While both compounds present a strong anisotropy at sufficient small doping, the main difference lies in the level of doping required to make the anisotropy negligble. While for \mbtl\, this occurs at small doping ($\mu\sim50$meV, $n\sim5\times10^{19}$\,electron/cm$^3$), for \mbt\, the anisotropy is preserved in a broader range of doping, which includes all estimates based on Hall-data reported in the literature.

\begin{figure}[h]
	\centering
	\includegraphics[width=7.5cm,angle=0]{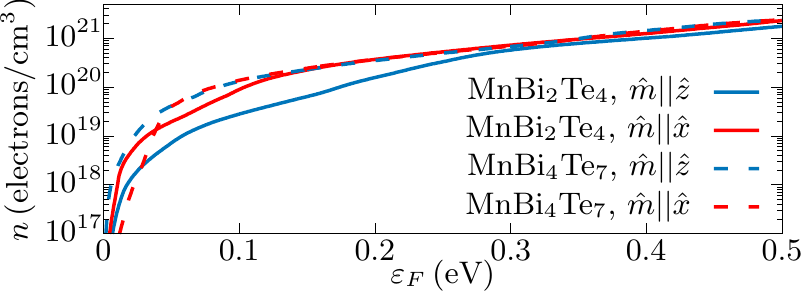}
	\caption{Carrier density $n$ vs Fermi energy $\varepsilon_F$ for \mbt\, and \mbtl, for different orientations of the magnetic moments.
	}
	\label{fig:dft3} 
\end{figure}

\clearpage

\end{document}